\documentclass[twocolumn]{aastex62}

\usepackage{xcolor}
\usepackage{amsmath}
\usepackage{graphicx}
\usepackage[normalem]{ulem}

\received{\today}

\shorttitle{Intermittent Temperature-Anisotropy}

\begin{document}

	\title{Intermittency and Ion Temperature-Anisotropy Instabilities: Simulation and Magnetosheath Observation}

	\author[0000-0001-8358-0482]{Ramiz~A. Qudsi}
	\affiliation{Department of Physics and Astronomy, University of Delaware, Newark, DE 19716, USA}

	\author[0000-0002-6962-0959]{Riddhi Bandyopadhyay}
	\affiliation{Department of Physics and Astronomy, University of Delaware, Newark, DE 19716, USA}

	\author[0000-0002-2229-5618]{Bennett~A. Maruca}
	\affiliation{Department of Physics and Astronomy, University of Delaware, Newark, DE 19716, USA}
	\affiliation{Bartol Research Institute, Newark, DE, USA}

	\author[0000-0003-0602-8381]{Tulasi~N. Parashar}
	\affiliation{Department of Physics and Astronomy, University of Delaware, Newark, DE 19716, USA}
	\affiliation{School of Chemical and Physical Science, Victoria University of Wellington, Kelburn, Wellington 6012, NZ }

	\author[0000-0001-7224-6024]{William~H. Matthaeus}
	\affiliation{Department of Physics and Astronomy, University of Delaware, Newark, DE 19716, USA}
	\affiliation{Bartol Research Institute, Newark, DE, USA}

	\author[0000-0001-8478-5797]{Alexandros Chasapis}
	\affiliation{Laboratory for Atmospheric and Space Physics, University of Colorado Boulder, Boulder, Colorado, USA}

	\author[0000-0002-4655-2316]{S. Peter Gary}
	\affiliation{Space Science Institute, Boulder, CO, USA}

	\author[0000-0001-8054-825X]{Barbara~L. Giles}
	\affiliation{NASA Goddard Space Flight Center, Greenbelt, Maryland 20771, USA}

	\author[0000-0003-1304-4769]{Daniel~J. Gershman}
	\affiliation{NASA Goddard Space Flight Center, Greenbelt, Maryland 20771, USA}

	\author[0000-0001-9228-6605]{Craig~J. Pollock}
	\affiliation{Denali Scientific, Fairbanks, Alaska 99709, USA}

	\author[0000-0001-9839-1828]{Robert~J. Strangeway}
	\affiliation{University of California, Los Angeles, California 90095-1567, USA}

	\author[0000-0001-7188-8690]{Roy~B. Torbert}
	\affiliation{University of New Hampshire, Durham, New Hampshire 03824, USA}

	\author[0000-0002-3150-1137]{Thomas~E. Moore} 
	\affiliation{NASA Goddard Space Flight Center, Greenbelt, Maryland 20771, USA}

	\author[0000-0003-0452-8403]{James~L. Burch}
	\affiliation{Southwest Research Institute, San Antonio, Texas 78238-5166, USA}

		
		
	\begin{abstract}
		Weakly collisional space plasmas are rarely in local thermal equilibrium and often exhibit non-Maxwellian electron and ion velocity distributions that lead to the growth of microinstabilities, that is, enhanced electric and magnetic fields at relatively short wavelengths. These instabilities play an active role in the evolution of space plasmas, as does ubiquitous broadband turbulence induced by turbulent structures. This study compares certain properties of a 2.5 dimensional Particle-In-Cell (PIC) simulation for the forward cascade of Alf\'venic turbulence in a collisionless plasma against the same properties of turbulence observed by the Magnetospheric Multiscale Mission spacecraft in the terrestrial magnetosheath. The PIC simulation is of decaying turbulence which develops both coherent structures and anisotropic ion velocity distributions with the potential to drive kinetic scale instabilities. The uniform background magnetic field points perpendicular to the plane of the simulation. Growth rates are computed from linear theory using the ion temperature anisotropies and ion beta values for both the simulation and the observations. Both the simulation and the observations show that strong anisotropies and growth rates occur highly intermittently in the plasma, and the simulation further shows that such anisotropies preferentially occur near current sheets. This suggests that, though microinstabilities may affect the plasma globally , they act locally and develop in response to extreme temperature anisotropies generated by turbulent structures. Further studies will be necessary to understand why there is an apparent correlation between linear instability theory and strongly intermittent turbulence.

		\keywords{plasmas --- turbulence --microinstabilities}

	\end{abstract}


	\section{Introduction} \label{sec:intro}
	
		The magnetized plasma of the solar wind is heated and accelerated in the solar corona, from which it flows continuously and supersonically into deep space. Earth's magnetic field obstructs and deflects part of this flow, which results in a region of subsonic solar plasma known as the magnetosheath.

		The low density and extreme dynamics of space plasmas such as these ensure that they almost invariably deviate substantially from local thermal equilibrium \citep{Marsch2006a,Verscharen2019}. For example, even though the majority of solar-wind ions are protons (ionized hydrogen) or $\alpha$-particles (fully ionized helium), these two particle species rarely have equal temperatures or bulk velocities \citep[see, e.g.,][]{Feldman1974a,Marsch1982,Hefti1998,Kasper2008,Maruca2013b}. Furthermore, the velocity distribution function (VDF) of any given ion species often significantly departs from the entropically favored Maxwellian condition functional form \citep{Feldman1973a,Feldman1974,Marsch1982b,Alterman2018}.
	
		The study described herein focuses on protons and in particular on their temperature-anisotropy: a phenomenon where the VDF lacks spherical symmetry. Observations of solar wind and magnetosheath from multiple spacecraft \citep{Feldman1973b,Marsch1982b,Kasper2002} have shown that the core of proton VDFs can often be approximated as ellipsoidal and aligned with the plasma's local magnetic field, $\mathbf{B}_0$. Consequently, the protons exhibit distinct kinetic temperatures, $T_{\perp p}$ and $T_{\parallel p}$, in the directions perpendicular and parallel to $\mathbf{B_0}$. Proton temperature anisotropy is commonly quantified by the ratio ($R_p \equiv T_{\perp p}\,/\,T_{\parallel p}$). Both values of $R_p > 1$ and $R_p < 1$ are commonly observed in the solar wind and in Earth's magnetosheath.
	
		If $R_p$ departs sufficiently from unity, it can trigger a kinetic microinstability: a short-wavelength fluctuation with an exponentially growing amplitude that is fueled by the VDF's free energy (see Section~\ref{sec:bgnd}). The threshold $R_p$-value for the onset of a proton temperature-anisotropy instability depends on all plasma parameters (e.g., composition and relative temperatures) but depends most strongly on proton parallel beta,
		\begin{equation} \label{eqn:betaparp}
			\beta_{\parallel p} = \frac{n_p\,k_B\,T_{\parallel p}}{B_0^2\,/\,(2\,\mu_0)} \ ,
		\end{equation}
		where $n_p$ is the proton number density, $k_B$ is the Boltzmann constant, and $\mu_0$ is the permeability of free space. A detailed and comprehensive discussion on linear growth rates, calculation methodology, and threshold values can be found in \citet{Gary1993}.

		These instabilities have threshold $R_p$-values, which means that they can effectively limit the degree to which proton temperature can depart from isotropy. If an unstable mode grows and does not saturate, it eventually becomes nonlinear, scatters particles in phase space, and drives the VDF toward local thermal equilibrium. Multiple studies have analyzed large datasets from various spacecrafts and under the assumptions of a spatially homogeneous plasma and a bi-Maxwellian proton velocity distribution, such studies have found that the joint distribution of $(\beta_{\parallel p},R_p)$-values from the interplanetary solar wind largely conforms to the limits set by the instability thresholds \citep{Gary2001,Kasper2002,Hellinger2006,Matteini2007}. A recent study by \citet{Maruca2018} confirmed the same effect in Earth's magnetosheath. Additional studies have found that plasma with unstable $(\beta_{\parallel p},R_p)$-values is statistically more likely to exhibit enhancements in magnetic fluctuations \citep{Bale2009} and proton temperature \citep{Maruca2011}. These findings suggest that the instabilities not only regulate temperature anisotropy in space plasmas but, in doing so, play an integral role in the large-scale evolution of the plasmas.
	
		Despite these extensive statistical studies, relatively little attention has been devoted to understanding the spatial distribution of the unstable modes within the plasma. The empirical studies of $(\beta_{\parallel p},R_p)$-distributions --- especially that by \citet{Matteini2007} --- indicate that the instabilities 		globally limit proton temperature anisotropy and affect the large-scale 		thermodynamics of expanding solar-wind plasma. Nevertheless, the instabilities themselves act on far smaller scales. Indeed, \cite{Osman2012} found that unstable $(\beta_{\parallel p},R_p)$-values are statistically more likely to exhibit enhanced values of the 	partial variance of increments (PVI), which is an indicator of intermittent structure. This result suggests that long-wavelength turbulence may play a substantial role in generating the local plasma conditions that drive these microinstabilities. Also, advancements made in numerical simulation by \citet{Servidio2012a, Greco2012, Servidio2015}, with corroboration from space plasma observations \citep{Marsch1992, Sorriso-Valvo1999, Osman2011, Osman2012, Kiyani2009} shows the importance of intermittency in interpretation of these observations.

        The question of where these instabilities develop raises the more fundamental issue of reconciling the assumptions of theory of microinstabilites, which assumes a homogeneous background, with the observed state of space plasma, which is rarely homogeneous. In the theory of microinstability, instability thresholds are computed using linear Vlasov theory. However, multiple studies have shown space plasma to be highly structured and thus inhomogeneous \citep{Burlaga1968, Tsurutani1979, Ness2001, Osman2012, Osman2012a, Greco2012}. This has been a persistent question in the field and a prime motivation for our work which formalizes the implications of \citet{Osman2012}.

        Analyses of both high-resolution kinetic simulation and high-cadence in-situ observations reveal that proton temperature-anisotropy instabilities are distributed intermittently in physical space. Section~\ref{sec:bgnd} lays the background of linear theory and instability threshold calculations and discusses the underlying inconsistencies. The computed growth rates were used to interpret both the kinetic simulation and in-situ observations, which are discussed in Sections~\ref{sec:pic} and~\ref{sec:mms}, respectively. Section~\ref{sec:con} summarizes the conclusions of this study.

    \section{Background} \label{sec:bgnd}

        \subsection{Microinstabilities in plasma} \label{subsec:imhgn}

            Linear growth rates are calculated using the linear dispersion relation obtained from Vlasov equation. \citet{Gary1993} discusses the computation methodology in great detail. We use the value of these growth rates to determine if the plasma VDF in a certain space at a certain time is susceptible to being unstable or not. For this purpose we set a cut-off for $\gamma_{jp}/\Omega_p$ at $10^{-5}$, where $\gamma_{jp}$ is one of the four microinstability growth rates arising because of proton temperature-anisotropy and $\Omega_p$ is the proton cyclotron frequency. For parallel propagation, we get cyclotron and parallel firehose instability for $R_p > 1$ and $R_p < 1$ respectively. The oblique (non-propagating) modes are mirror for $R_p > 1$ and oblique firehose for $R_p < 1$.

            Though linear theory works well for plasma with homogeneous background, when it comes to its application to space plasmas to study the characteristics of space plasmas, the method is not without caveats. Multiple studies have shown space plasma to be highly structured and thus inhomogeneous \citep{Burlaga1968, Tsurutani1979, Ness2001, Osman2012, Osman2012a, Greco2012}. In fact, by all accounts inhomogeneity is ubiquitously present in the space plasma, and thus any study of instabilities in plasma should take into account the inhomogeneity of the background among variation in other parameters.

            Consequently, use of linear theory for such studies of course presents a theoretical inconsistency in the application of computed instability thresholds to study the properties of plasma because of the underlying disparity between the assumptions of linear theory and observed space plasma. However, several studies over the last three decades have presented empirical evidence of agreement between the observations and theoretical prediction \citep{Gary1991, Gary1994, Gary2001, Gary2006, Kasper2002, Hellinger2006, Maruca2011, Maruca2012, Maruca2018}. These studies strongly suggest that linear instability thresholds are indeed efficient in restricting the plasma/plasma VDF in a narrow region of $\beta_{\parallel p}-R_p$ plane inhibiting excursion of plasma VDFs to extreme anisotropy regions at high $\beta_{\parallel p}$. Although limitations on spatial and temporal resolution using present-day spacecraft make it difficult to directly demonstrate the existence of such instabilities in space plasmas, work done by, \citet{Bale2009, He2011, Podesta2013, Jian2009, Jian2010, Jian2014, Klein2014, Telloni2016, Gary2016} and others provide indirect evidence for the presence of various different instabilities. More details could be found in \citet{Verscharen2019} and references therein.

            An ideal study would indeed include the effect of background inhomogenities in computing the growth rates. However we do not have any such established methodology and development of such a method is beyond the scope of this study. We thus are restricted to use the established theory of microinstabilities, and calculate instability thresholds from linear Vlasov equations. Although we do not discuss the consequences of electron anisotropies here, we note that both computer simulations and magnetosheath observations \citep{Gary2005} have shown that electron temperature anisotropies in collisionless plasmas can drive whistler instabilities which, in turn, scatter the electrons to establish a constraint on the anisotropy of that species \citep{Gary1996}, in full analogy with the case of ion instabilities and anisotropy constraints discussed here.

	\section{Results: PIC simulation} \label{sec:pic}
	
		We first applied our linear Vlasov calculations to the output of fully kinetic, particle-in-cell (PIC) simulation in homogeneous, collisionless, magnetized plasmas with $\mathbf{B}_0$ in the $z$-direction that we implemented with the P3D code \citep{Zeiler2002}. As this was a 2.5D simulation, all vector quantities were modeled as having three components, but all plasma parameters varied only in the $xy$-plane. For this study the initial conditions were chosen such that the particle distribution was Maxwellian, $\beta_p=\beta_e=\, 1.2, R_p=1, T_p = T_e$ and the rms value of fluctuations in magnetic and velocity fields were half of the background values. High $\beta$ values as well as values much lower than 1 makes the PIC computations very expensive and thus were avoided. The system was then allowed to evolve without any external forcing. Fluctuation in the observed magnetic and velocity fields produce and drive the turbulence in the plasma. All the analyses presented here, are performed near the instant at which the mean-square of out of plane current $(J_z)$ peaks, when the non-linear processes are known to be most active. More details about the simulation can be found in \citet{Parashar2018}.

\begin{figure*}
			\begin{center}
				\includegraphics[width=1.\textwidth]{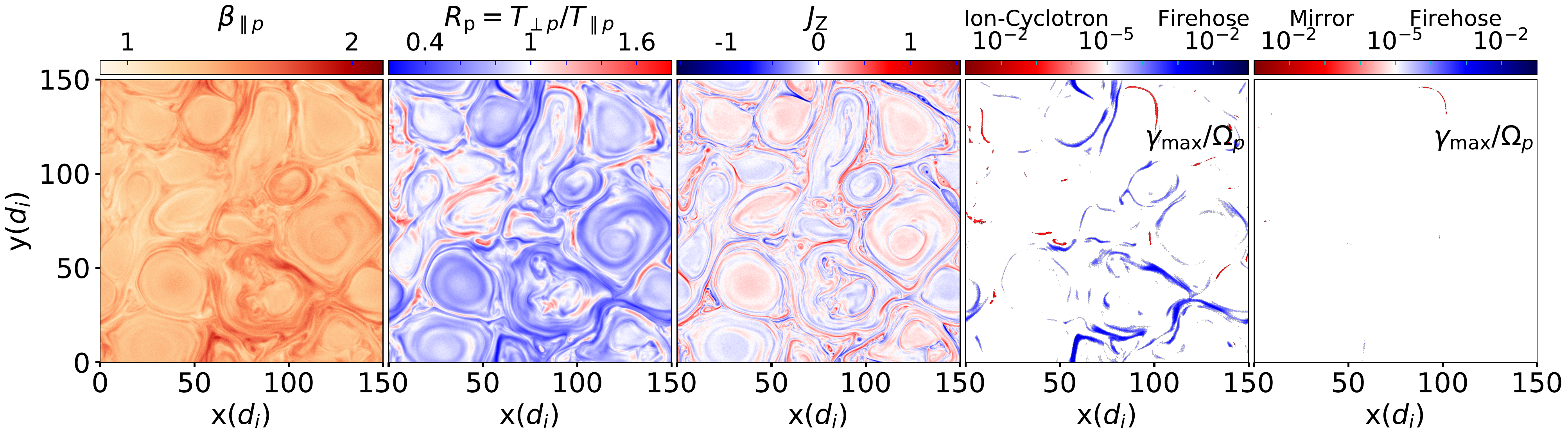}
				\caption{Colorplot of (left to right) $\beta_{\parallel p}, R_p \mathrm{\,and\,} J_z$ from a fully kinetic 2.5D PIC simulation with the initial condition of $\beta_p=\beta_e = 1.2, T_p=T_e \,\mathrm{and}\, R_p=1$. Fourth and fifth panels show the spatial distribution of $\gamma_{\max}$ for parallel and oblique propagation respectively corresponding to first two panels}
				\label{fig:brj}
			\end{center}
		\end{figure*}

		The first three panels of Figure $\ref{fig:brj}$ show the three parameters --- $R_p$, $\beta_{\parallel p}$, and $J_z$ --- across the simulation box. The system is strongly turbulent and exhibits structures of various scales. The extreme values of each parameter occur in distinct regions that occupy only small fractions of the total volume. That is, these quantities are intermittent, which is correlated with the existence of sharp gradients and coherent structures \citep{Matthaeus2015}. Further, the extreme values of $R_p$ and $\beta_{\parallel p}$ reside near (but not exactly coincident with) the extreme values of $J_z$. These concentrations of current densities frequently correspond to current sheets, as reported in \citet{Parashar2016}.

		Using the method described in Section \ref{sec:bgnd}, we computed $\gamma_\mathrm{max}$ for the $(\beta_{\parallel p},R_p)$-pair at each grid point of the simulation, where $\gamma_\mathrm{max}$ is the maximum value of growth rate for all possible values of propagation vector (\textbf{k}). The fourth and fifth panels of Figure~$\ref{fig:brj}$ show the spatial distribution of growth rates for the solutions with positive growth rates, corresponding to the first two panels of the same figure. As described in Section \ref{sec:bgnd}, for $\gamma_\mathrm{max}$, we imposed a cut-off at $10^{-5} \Omega_p$; thus growth rates less than $10^{-5} \Omega_p$ are considered to be 0. The fourth panel of Figure~$\ref{fig:brj}$ corresponds to the parallel modes (cyclotron for $R_p\textgreater \,1$ and parallel firehose for $R_p\textless \,1$), whereas the fifth panel is for the oblique propagation (mirror for $R_p\textgreater \,1$ and oblique firehose for $R_p\textless \,1$). The paucity of blue color in the fifth panel implies that the $\beta_{\parallel p}$ (and/or $R_p$) was rarely high (low) enough to excite any mode of oblique firehose instability.

		Comparing the second panel to the fourth and fifth of Figure \ref{fig:brj}, we see that values of $\gamma_\mathrm{max}>0$ are concentrated in distinct, filament-like regions of the $xy$-plane where extreme values of temperature anisotropy also occur. We note that, because the simulation is 2.5D with $\mathbf{B_0}$ perpendicular to the simulation plane, the growth of instabilities such as the proton cyclotron and the parallel proton firehose with maximum growth at $\mathbf{k \times B_0 = 0}$ is suppressed.

	\section{Results: MMS Observations} \label{sec:mms}
	
		We carried out a similar analysis on an interval of burst-mode measurements of the magnetosheath from Magnetospheric Multiscale (MMS).

		MMS is a constellation of four identical spacecraft designed to study reconnection in the magnetosphere of the Earth \citep{Burch2016}. We used proton density and temperature-anisotropy data from the Fast Plasma Investigation (FPI) and magnetic-field data from the Fluxgate Magnetometer (FGM). In burst mode FPI measures one proton distribution every 150 ms \citep{Pollock2016}, and the cadence of FGM is 128 Hz~ \citep{Russell2016}.

		Using the measured temperature-anisotropy and magnetic field vectors, we computed the value of the linear instability growth rates ($\gamma_{\max}$) for each point in the time series using the same methodology as described in Section \ref{sec:pic}. For this analysis we chose a 40-minute long period of burst data from 26-12-2017 starting at 06:12:43 UTC. This period was chosen in part because of its relatively long duration compared to typical burst mode intervals. During this period average proton density was and $22 \,\mathrm{cm}^{-3}$, the average value of $\beta_{\parallel p} \mathrm{was}$ 4.5 and average bulk velocity of the plasma was $238\, \mathrm{km/s}$. More details about this particular period can be found in \citet{Parashar2018a}.
		\begin{figure}
			\begin{center}
				\includegraphics[width=0.46\textwidth]{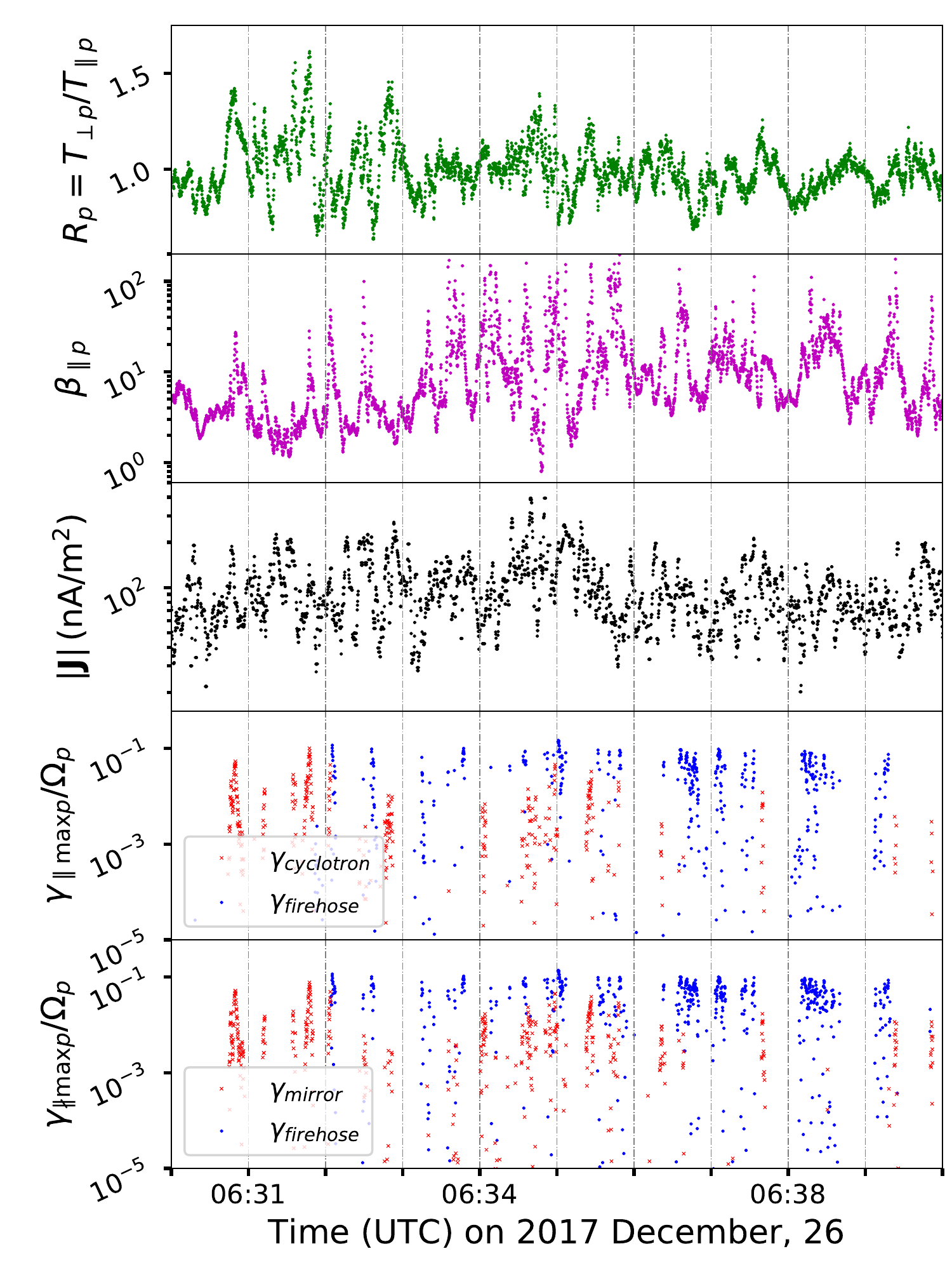}
				\caption{Time series plot of proton anisotropy ratio $(R_p)$, proton parallel beta $(\beta_{\parallel p})$, total current density in $\mathrm{nA/m^2}$, parallel instability growth rates (proton-cyclotron, in red and parallel firehose in blue), and oblique instability growth rates (mirror in red and oblique firehose in blue), as observed by MMS on 2017 December 26}
				\label{fig:mms_all}
			\end{center}
		\end{figure}
	
		Figure $\ref{fig:mms_all}$ shows a typical 10 minute portion of the time series of the data discussed above. The panels, from top to bottom show $R_p, \beta_{\parallel p}, |\mathbf{J}|$ and maximum growth rates ($\gamma_\mathrm{max}$) for parallel and oblique instabilities respectively.

        \begin{figure}
			\begin{center}
				\includegraphics[width=0.50\textwidth]{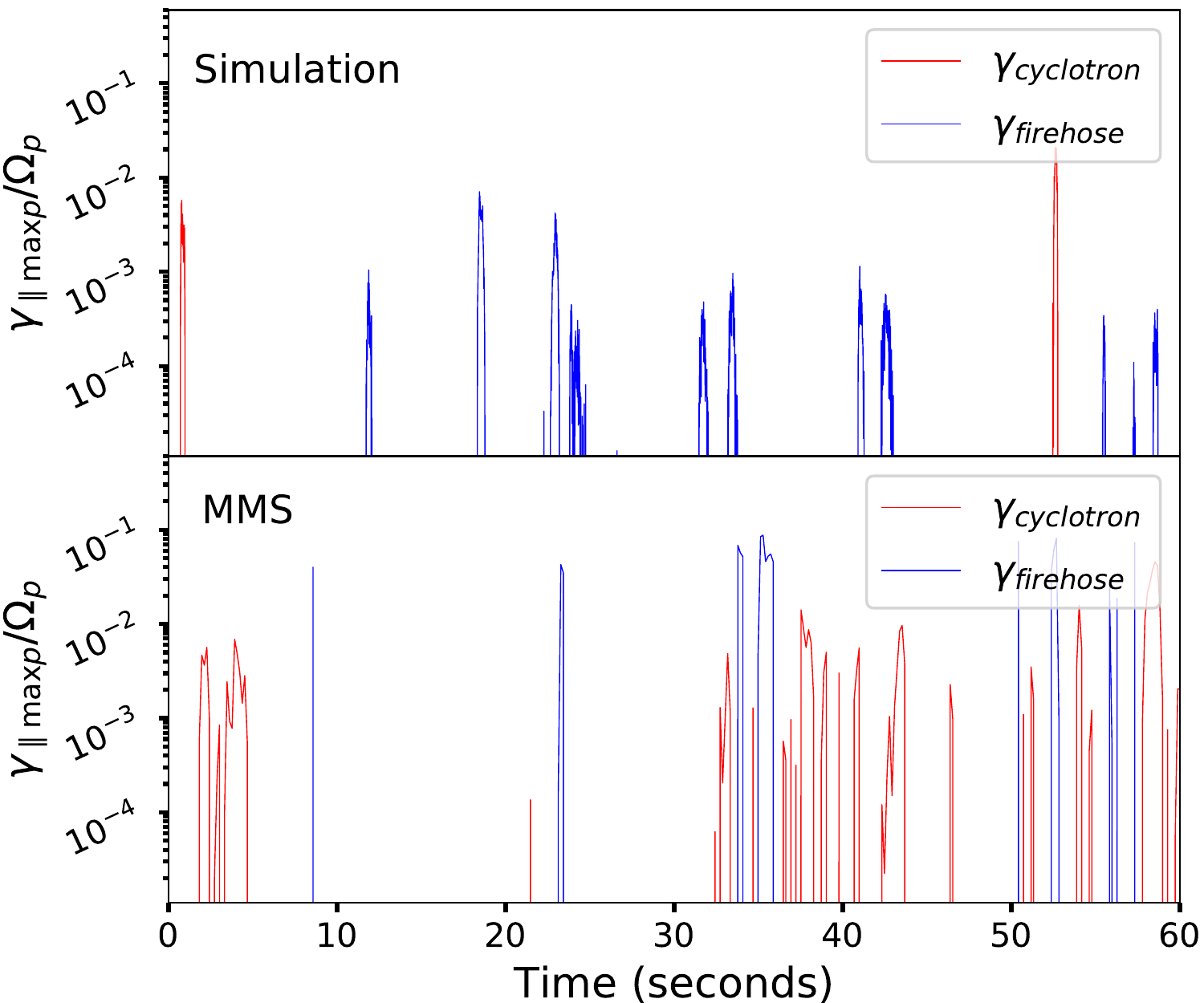}
				\caption{Comparison of simulation and MMS time series for $\gamma_{\parallel max}$ values for a 1-minute period. The top panel shows the distribution for a 1-minute long flight through the simulation box and the lower panel shows the distribution of $\gamma_\mathrm{max}$ starting at 06:34 on 2017 December 26}
				\label{fig:spc}
			\end{center}
		\end{figure}

		Comparing Figures \ref{fig:brj} and \ref{fig:mms_all} we see that a larger 
		fraction the MMS data ($30\%$) are unstable versus grid points from the simulation
		($0.8\%$), with $\gamma_\mathrm{max}$ values above the cut-off
		($10^{-5}\Omega_p$). This discrepancy arises in part because MMS data have much 
		higher values of $\beta_{\parallel p}$ than the simulation (median values of 4.5 and 1.2, respectively). 
		 Furthermore \citet{Servidio2015} found that, for a given value 
		of $\beta_{\parallel p}$, simulations of the turbulence 
		type, like in the present case, generally admit less extreme temperature-anisotropy 
		than are seen in space observations, because typical simulations are of modest size and lack large scale driving.

		The time series for MMS observation (Figure $\ref{fig:mms_all}$) exhibits 
		intermittent structure in the distribution of growth rates that are 
		similar to what we see in panel 4 and 5 of Figure $\ref{fig:brj}$ for 
		simulation. Figure~\ref{fig:spc}, which shows the comparison of the time 
		series of simulation and MMS data for a 1-minute period, shows that 
		qualitatively they have similar distribution. Time series for the 
		simulation was computed by flying a virtual spacecraft, travelling 
		at the plasma bulk speed (238 km/s), through the 
		entire box at an angle of 75 degrees with respect to x-direction.

		In Figure \ref{fig:mms_all} the points of instabilities ($\gamma_\mathrm{max}>0$) 
		are concentrated together, spreading over a small time interval lasting 
		typically a few seconds (4-8 seconds) with sharp peaks. 
		Though in this study we did not quantify the length scale of all the peaks, we found that typically they are spread over a length scale of 
		$\sim 20-40 \mathrm{d_i}$, where $\mathrm{d_i}$ is the ion-inertial length and the length scale was calculated using 
		the flow speed of the plasma and the duration of the peak.

	\section{Discussion and Conclusion} \label{sec:con}

		In recent years, two different perspectives have been widely used to explain the behavior of the solar wind, magnetosheath, and similar space plasmas. In the first picture, the linear theory of plasma instability, at high $\beta_{\parallel p}$, for extreme $R_p$, different instability thresholds become active, thereby confining the plasma population to lower values of $R_p$ \citep{Gary2001,Kasper2002,Hellinger2006,Matteini2007,Klein2018}. In the second, turbulence generates sharp gradients in the plasma that produce temperature anisotropy \citep{Osman2011,Greco2012,Valentini2014,Parashar2016}.

		These two theories have been non-reconcilable because of the basic underlying assumption. The linear theory of plasma instability assumes a homogeneous background magnetic field whereas turbulence relies on large fluctuations in the field. It was hitherto unclear if these two seemingly disparate processes--microinstabilities and turbulence--are connected in any way in configuration space. The apparent contradiction---homogeneity against intermittent inhomogeneity---between the two interpretations poses a question of fundamental importance in the study of space plasmas specifically and collisionless plasmas in general:  How can an inhomogeneous phenomenon such as turbulence be consistent with anisotropy constraints derived from linear theory of homogeneous plasmas? Our simulation shows that the turbulence indeed heats the plasma anisotropically. But the simulation also show that these anisotropies are strongly localized; furthermore the 2.5D character of the simulation with a strong background magnetic field out of the simulation plane acts against the growth of the proton cyclotron and parallel proton firehose microinstabilities which are the more likely sources of the proton anisotropy constraints.  So it appears that the best we can say now is that at present we do not understand why there should be a correlation between linear theory and strongly intermittent turbulence.  Clearly, further studies are necessary to resolve this apparent contradiction.

		In Figure \ref{fig:brj} the regions of significant growth rates are close to the regions of peak current values. This suggests that current sheets are producing the extreme temperature-ansiotropies that ultimately drive the instabilities. Note, though, that the high-$\gamma_{\max}$ regions and the high-$J_z$ regions do not perfectly overlap: they tend to be adjacent to each other rather than co-located, as seen in \citet{Greco2012}. Thus, traditional methods of correlation calculation would be inadequate to quantify the relationship between these two structures. Instead, an analysis using cross correlations of these quantities (see e.g. \cite{Parashar2016}) or joint distributions (see e.g. \cite{Yang2017}) to explore the causal connection between instabilities and turbulence-generated current sheets would be the next step forward.
		
		In this paper we have found an explicit connection between intermittency in plasma turbulence, and the local enhancement of linear instability growth rates. Intermittency, or burstiness, in measured properties of turbulence is typically associated with the dynamical formation of coherent structures in space. In hydrodynamics these structures would be found in the vorticity or velocity increments. In the magnetohydrodynamics or plasma context, intermittent structures of other types can be found, such as sheets or cores of electric current density \citep{Carbone1990, Biskamp1986}. A phenomenology of intermittent plasma structures has begun largely due to advances in numerical simulation (e.g. \citet{Servidio2012a, Greco2012, Servidio2015} ) with confirmation and guidance emerging from observations in space \citep{Marsch1992, Sorriso-Valvo1999, Horbury1997, Osman2011, Osman2012, Kiyani2009}. Intermittency is clearly influential in the interpretation of observations, while its theoretical importance derives from its potential connection to the nature and statistics of dissipation \citet{Kolmogorov1962, Karimabadi2013, Wan2016, Howes2015, Matthaeus2015}.  The connection we have found here -- that linear instability growth rates computed from (admittedly oversimplified) homogeneous plasma theory, also occur in intermittent bursts -- adds to this emerging understanding of plasma dissipation. Previous studies found that pathways, such as inertial range transfer \citep{SorrisoValvo2019}, electromagnetic work \citep{Wan2012}, electron energization \citep{Karimabadi2013}, and pressure-strain interactions \citep{Yang2017} concentrate in small sub-volumes of plasma turbulence. Dynamical processes that lead to dissipation such as magnetic reconnection, also occur in spatially localised regions \citep{Drake2008}. Along with these we now have seen that velocity-space driven phenomena \citep{Servidio2012a, Greco2012, Schekochihin2016, Servidio2015} also occur in similar highly localized sub-volumes. The nature of the spatial or regional correlations of these kinetic processes to the surrounding dynamical processes that drive them largely remains to be explored.

	\section*{Acknowledgments}
	
	    T.N.P. was supported by NSF SHINE Grant AGS-1460130 and NASA HGI Grant 80NSSC19K0284. WHM is a member of the MMS Theory and Modeling Team and was supported by NASA Grant NNX14AC39G. The research of SPG was supported by NASA grant NNX17AH87G.\\
	    This study used Level 2 FPI and FIELDS data according to the guidelines set forth by the MMS instrumentation team. All data are freely available at https://lasp.colorado.edu/mms/sdc/.\\
	    We thank the MMS SDC, FPI, and FIELDS teams for their assistance with this study. We acknowledge high-performance computing support from Cheyenne (doi:10.5065/D6RX99HX) provided by NCAR’s Computational and Information Systems Laboratory, sponsored by the National Science Foundation. These simulations were performed as part of the Accelerated Scientific Discovery Program (ASD).\\
	    The author would like to thank Rohit Chibber and Mark Pultrone for their inputs on the draft and several useful discussions.

\bibliographystyle{aasjournal}

\end{document}